# Competition between Weak Localization and Antilocalization of Dirac-like Fermions in a Spin-Polarized Two-Dimensional Electron Gas at KTaO$_3$ (111) Interface

Hui Zhang,[1,2]* Daming Tian,[1] Xiaobing Chen,[3]* Lu Chen,[1] Min Li,[1] Yetong Bai,[1] Fengxia Hu,[4,5,6] Baogen Shen,[4,5,6,7] Jirong Sun,[4,8]* and Weisheng Zhao[1,2]*

[1]School of Integrated Circuit Science and Engineering, Beihang University, Beijing 100191, China

[2]State Key Laboratory of Spintronics Hangzhou International Innovation Institute, Beihang University, Hangzhou 311115, China

[3]Quantum Science Center of Guangdong-Hong Kong-Macao Greater Bay Area (Guangdong), Shenzhen 518045, China

[4]Beijing National Laboratory for Condensed Matter Physics, Institute of Physics, Chinese Academy of Sciences, Beijing 100190, China

[5]School of Physical Sciences, University of Chinese Academy of Sciences, Beijing 100049, China

[6]Songshan Lake Materials Laboratory, Dongguan, Guangdong 523808, China

[7]Ningbo Institute of Materials Technology & Engineering, Chinese Academy of Sciences, Ningbo, Zhejiang, 315201, China

[8]School of Physics, Zhejiang University, Hangzhou 310027, China

*Corresponding authors:
huizh@buaa.edu.cn
chenxiaobing@quantumsc.cn
jrsun@iphy.ac.cn
wszhao@buaa.edu.cn

**Abstract**

Quantum transport phenomena in two-dimensional electron gases (2DEGs) at oxide interfaces have garnered significant interest owing to their potential in spintronic and quantum information technologies. Here, we systematically investigate the quantum conductance corrections of spin-polarized 2DEGs formed at the interfaces between two insulating oxides, ferromagnetic $EuTiO_3$ (ETO) films and (111)-oriented $KTaO_3$ (KTO) substrates. The anomalous Hall effect and hysteretic magnetoresistance provide clear evidence for long-range ferromagnetic order in the 2DEGs, which could be attributed to interfacial Eu doping in combination with the magnetic proximity effect of the ETO layer. The breaking of time-reversal symmetry by ferromagnetism in the 2DEGs, and with the assistance of spin-orbit coupling effect, gives rise to a nontrivial Berry phase. This results in a competition between weak localization (WL) and weak antilocalization (WAL) in the quantum transport of Dirac-like fermions at the KTO (111) interfaces. Notably, this competitive behavior can be effectively tuned by optical gating via a photoexcitation-induced shift of the Fermi level. Our findings demonstrate a controllable platform based on spin-polarized oxide 2DEGs for quantum transport, opening new avenues for spin-orbitronic and topological electronic applications.

**Introduction**

Quantum interference effects in low-dimensional electron systems have attracted extensive attention, as they provide an effective means to probe phase coherence, spin-orbit coupling (SOC), and disorder[1-7]. The resulting quantum corrections to conductivity, manifested as weak localization (WL) or weak antilocalization (WAL), can be detected via measuring the magnetoconductivity (MC) under perpendicular magnetic fields[7]. Two-dimensional electron gases (2DEGs) formed at oxide interfaces have emerged as a versatile platform for exploring quantum phenomena, owing to broken spatial inversion symmetry, strong electron correlations, and Rashba-typed SOC. In particular, magnetic 2DEGs offer unique opportunities for exploring the impact of time-reversal symmetry (TRS) breaking, which gives rise to a range of intriguing physical properties. Recently, anomalous quantum corrections to the MC of Dirac-like fermions have been observed in a spin-polarized oxide 2DEG at the (111)-oriented LaAlO$_3$/EuTiO$_3$/SrTiO$_3$ (LAO/ETO/STO) heterostructure[8]. The coexistence of Rashba SOC, ferromagnetic order, and hexagonal band warping at the (111) interface induces a nontrivial Berry phase, which manifests as competing WL and WAL scattering channels of Dirac-like fermions. Such behavior exhibits a phenomenology analogous to that of magnetically doped three-dimensional topological insulators (TIs)[9-13], including Cr-doped Bi$_2$Se$_3$[11], Mn-doped Bi$_2$Te$_3$[13] and Sb-doped Bi$_2$Te$_3$[12].

Beyond the 3d transition metal oxide STO, KTaO$_3$ (KTO)-based heterostructures also host 2DEGs originating from Ta 5$d$ orbitals[1,14,15]. A distinguishing feature of KTO 2DEGs is their large Rashba SOC, as revealed by WAL measurements, with the Rashba coefficient ($\alpha_R$) found to be 5~10 times larger than that of STO 2DEGs. This stronger SOC in KTO 2DEGs induces a more pronounced spin splitting in the band structure compared with STO counterparts[1,14-16]. In addition, spin-polarized 2DEGs at KTO interfaces have been realized via magnetic proximity effects, as demonstrated in EuO/KTO[17,18] and EuO/LaTiO$_3$/KTO systems[19]. Importantly, the combination of broken time-reversal and spatial inversion symmetries together with strong SOC makes spin-polarized 2DEGs at (111)-oriented KTO interfaces an ideal platform for exploring topologically non-trivial states. The buckled honeycomb lattice of the KTO (111) surface can host exotic band structures, such as Dirac-like dispersions. However, quantum interference of Dirac-like fermions in spin-polarized 2DEGs at KTO (111) interfaces has not yet been reported, despite its significance in elucidating the interplay among magnetism, SOC, and topologically non-trivial states in the KTO system.

In this work, we systematically investigate the quantum conductance corrections of spin-polarized 2DEGs at the ETO/KTO (111) interfaces. Long-range ferromagnetic order in the 2DEG is confirmed by the observation of the anomalous Hall effect (AHE) and hysteretic magnetoresistance (MR). Temperature-dependent magnetotransport analysis reveals a competition between WL and WAL of Dirac-like fermions, arising from time-reversal symmetry breaking caused by ferromagnetism in the 2DEGs below the magnetic transition temperature. Furthermore, the transport behavior of the 2DEG can be remarkably tuned by light illumination through the generation of a second species of highly mobile photoexcited carriers. Optical gating can also effectively modulate the competition between WL and WAL by shifting the Fermi level ($E_F$), indicating that the quantum interference is strongly dependent on band filling.

## RESULTS AND DISCUSSION
### 1. Structural characterization of the ETO/KTO (111) interface

ETO thin films were grown on (111)-oriented KTO single-crystal substrates using the pulsed laser deposition (PLD) technique (see the Methods section for details). Atomic force microscopy (AFM) characterization indicates that the 20-nm-thick ETO film exhibits an atomically flat surface morphology, with a root-mean-square (RMS) roughness of ~0.32 nm (Fig. S1). Bulk ETO crystallizes in a cubic structure with a lattice constant of 3.904 Å, which yields a small tensile lattice mismatch of ~2.1% with cubic KTO ($a_{KTO}$ = 3.989 Å), favorable for achieving a high-quality epitaxial growth. The formation of a single-crystalline ETO film is confirmed by X-ray diffraction (XRD) $\theta$-$2\theta$ scan and reciprocal space mapping (Fig. S2). To gain a deeper insight into the atomic-scale structure, high-resolution scanning transmission electron microscopy (STEM) was employed to examine the lattice structure of the ETO/KTO (111) heterostructure. A typical cross-sectional high-angle annular dark-field (HAADF)-STEM image of the ETO/KTO (111) interface, recorded along the [11-2] zone axis of KTO, is shown in Fig. 1a. A large-scale HAADF-STEM image is shown in Fig. S3. The ETO film exhibits high-quality epitaxial growth on the KTO substrate, forming a structurally coherent heterointerface without visible crystalline defects. Remarkably, a distinct interfacial layer with a thickness of ~1 nm is observed between ETO and KTO, as marked by the red dashed lines. The corresponding energy-dispersive X-ray spectroscopy (EDS) elemental mapping of the same region (Fig. 1b) further reveals the elemental composition and spatial distribution across the ETO/KTO (111) interface. Perovskite-typed KTO substrates are prone to losing A-site element K at high temperatures. During sample preparation, the volatilization of surface K facilitates the diffusion of Eu into the KTO substrate, resulting in an Eu-doped interfacial layer where Eu atoms substitute for K, as evidenced by the EDS mapping in Fig. 1b. In addition, Ti atoms also diffuse into the interfacial region, partially substituting for Ta. A high-magnification HAADF-STEM image near the ETO/KTO (111) interface is shown in Fig. 1c, overlaid with the corresponding atomic model. Figure 1d presents the atomic line profile extracted from the region marked by the yellow box in Fig. 1c, along the direction from substrate to film, revealing the intensity variations of the A- and B-site atoms across the interface. Notably, the K-site columns within the interfacial layer (indicated by pink arrows) exhibit higher intensities than those in the bulk, suggesting that some K atoms are substituted by heavier Eu atoms. Meanwhile, the Ta-site columns (indicated by purple arrows) display reduced intensities, indicating partial replacement by lighter Ti atoms. A combined analysis of the intensity profile and EDS elemental spectra confirms the presence of Eu, K, Ta, and Ti atoms within the interfacial layer. Similar interfacial elemental intermixing has also been observed at the ETO/KTO (001) heterointerface[20].

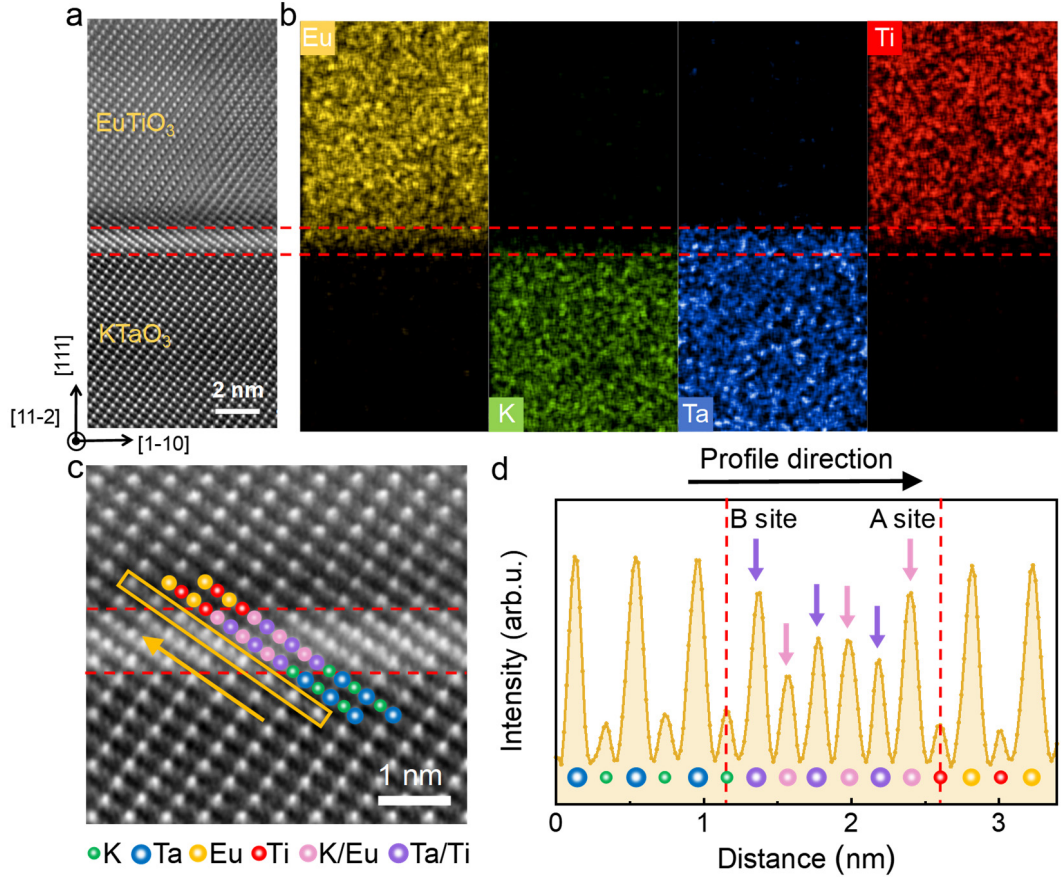

**Figure 1.** (a) Cross-sectional HAADF-STEM image of the interface between the ETO film and KTO substrate, recorded along the [11-2] zone axis. The red dashed lines indicate the interfacial layer. (b) Corresponding EDS elements mapping of the same region. (c) High-magnification HAADF-STEM image of the area near the ETO/KTO (111) interface shown in (a). (d) Atomic intensity line profile extracted from the region marked by the yellow box in (c), along the direction from substrate to film.

## 2. Competing WL and WAL in the magnetic 2DEG at the ETO/KTO (111) interface

The magnetic properties of the insulating ETO thin film were investigated and found to exhibit the standard ferromagnetic behavior characteristic of the strained ETO phase[21]. As shown in Fig. S4, the temperature-dependent magnetization (*M-T* curve) was measured under an in-plane magnetic field $B_\parallel$ = 0.05 T applied parallel to the film surface. A clear magnetic transition occurs at the Curie temperature (*Tc*) of ~7 K. The magnetic easy axis lies within the film plane (Fig. S4), with a saturation magnetization of ~6 $\mu_B$/Eu. These magnetic characteristics are in good agreement with previously reported epitaxial ETO films[18,20,22-25]. We further investigated the electronic transport properties of ETO ($t_{ETO}$)/KTO (111) samples with different ETO layer thicknesses ($t_{ETO}$ = 4, 8, and 20 nm). Although both ETO and KTO are insulators, all samples exhibit metallic behavior over a wide temperature range, indicating the formation of 2DEGs at the ETO/KTO (111) heterointerfaces (Fig. S5). Notably, a pronounced resistance upturn in the temperature-dependent sheet resistance ($R_S$-$T$ curve) emerges below ~20 K for all samples. This feature is generally attributed to the Kondo effect, which arises from carrier scattering by localized magnetic moments. To verify this, we fitted the low-temperature resistance upturn using the following equation[18,26]:

$$R_S(T) = R_0 + qT^2 + pT^5 + R_{K0}\left\{1 - \ln\left(\frac{T}{T_K}\right)\left[\ln^2\left(\frac{T}{T_K}\right) + S(S+1)\pi^2\right]^{-1/2}\right\} \quad (1)$$

where the first term is the residual resistance, the second term represents the Fermi liquid contribution, the third term accounts for electron-phonon interaction, and the last term corresponds to the Kondo effect. The Kondo term is described by the Hamann expression, where $R_{K0}$ is a temperature-independent resistance, $T_K$ is the Kondo temperature, and $S$ is the spin of the magnetic impurities. As shown in Fig. 2a, the normalized sheet resistance $R_S/R_{2K}$, where $R_{2K}$ is the sheet resistance at 2 K, fits well to Eq. (1), confirming the presence of the Kondo effect at the ETO/KTO (111) interface. The corresponding fitting parameters are summarized in Table S2. The extracted Kondo temperature $T_K$ increases monotonically from 4.1 to 6.6 K as $t_{ETO}$ increases from 4 to 20 nm, suggesting an enhancement of localized magnetic scattering. As supported by the STEM analysis discussed above, the diffusion of magnetic Eu atoms into the KTO substrate could induce localized magnetic moments, which may play a significant role in resulting in the Kondo effect. The increased ETO film thickness, which entails longer thermal exposure during growth, could promote elemental intermixing at the interface, thereby enhancing localized magnetic scattering and leading to a higher $T_K$.

Hall effect measurements show that the Hall resistance ($R_{xy}$) depends linearly on the out-of-plane magnetic field ($B_\perp$) with a negative slope, indicating that the charge carriers are electrons and only one type of carrier dominates the transport process. The temperature-dependent sheet carrier density ($n_S$) and Hall mobility ($\mu$) extracted from the Hall effect are shown in Fig. S6. To further investigate the anomalous Hall effect

(AHE), the anomalous Hall resistance ($R_{AHE}$) was obtained by subtracting the linear ordinary Hall background from the $R_{xy}$-$B_\perp$ curves. As shown in Fig. 2b, step-shaped $R_{AHE}$-$B_\perp$ curves are observed at 2 K for all samples, and the saturation $R_{AHE}$ increases from 0.64 to 1.15 Ω as $t_{ETO}$ increases from 4 to 20 nm. In addition to AHE, the temperature dependence of the magneticresistance (MR) was also investigated. Figure 2c presents the MR of the ETO (20 nm)/KTO (111) sample measured under an in-plane magnetic field $B_\parallel$ applied perpendicular to the current direction at various temperatures. When the magnetic field is cycled between ±0.3 T, hysteretic MR appears, providing a further evidence for the establishment of long-range magnetic order in the 2DEG. Notably, the temperature-dependent evolution of the MR-$B_\parallel$ loops exhibit a transition from butterfly-shaped hysteresis with twin peaks at low temperatures to twin-dip features at higher temperatures, accompanied by a crossover from negative to positive MR. The two neighboring extrema in the MR, located at ±0.01 T, are significantly larger than the coercive field of ETO thin films (±0.004 T) (Fig. S7). This behavior is markedly different from that of previously reported spin-polarized oxide 2DEGs, which originate from the magnetic proximity effect of EuO films[22,27]. In fact, the enhanced coercive field of the 2DEG can be attributed to magnetic atom doping[20], indicating that the ferromagnetism of the 2DEG may originate from both interfacial Eu doping and the magnetic proximity effect of the neighboring ETO layer. Upon further warming, the hysteretic MR behavior vanishes completely at 10 K. The above behavior is also observed when the in-plane magnetic field $B_\parallel$ is applied parallel to the current direction (see Fig. S8), indicating its isotropic nature, i.e., independence of the $B_\parallel$ direction. When the temperature is below the Kondo temperature $T_K$ = 6.6 K, the negative MR could be attributed to the Kondo effect, where conducting electrons are scattered by localized magnetic moments. Near the coercive field, where the magnetic configuration is most disordered, Kondo scattering and resistance reach their maximum. As the magnetic field aligns the randomly oriented magnetic moments, Kondo scattering is suppressed and the resistance is reduced, giving rise to twin resistance peaks in the MR loop[26]. In the high-temperature regime ($T$>$T_K$ = 6.6 K), the Kondo effect is not dominant. The MR exhibits two pronounced resistance dips at the coercive field, which is characteristic of long-range magnetic ordering[27].

The observations of the anomalous Hall effect and hysteretic MR loops confirm the formation of ferromagnetic 2DEGs at the ETO/KTO (111) interfaces. In such a spin-polarized 2DEG system, the interplay between strong SOC from the heavy 5$d$ element Ta and time-reversal-symmetry breaking due to ferromagnetism can give rise to exotic quantum states associated with the multiband structure of Ta 5$d$ orbitals. We investigate the effect of SOC on the electronic band structure of the 2DEG through first-principles calculation. The crystal structure of the (111)-oriented KTO-based oxide 2DEG can be regarded as a repetition of three inequivalent Ta atomic layers stacked along the [111] direction, forming a hexagonal lattice[8,28,29]. This geometric configuration, combined with spatial inversion symmetry breaking, lifts the threefold degeneracy of the Ta 5$d$ $t_{2g}$ orbitals via a trigonal crystal field under $C_{3v}$ symmetry, resulting in a non-degenerate $a_1$ state and a twofold-degenerate $e'$ doublet (see Fig. S9). When Rashba SOC is considered, the $e'$ orbitals further hybridize into states with angular momentum projections $L_{z,-}$ and $L_{z,+}$ (see Supplementary Note S1 for details). Figure 2d displays the electronic band structure of the 2DEG at the KTO (111) surface, calculated using density functional theory (DFT), with subbands characterized by $L_{z,+}$ (red) and $L_{z,-}$ (blue) orbitals. Notably, a pronounced Rashba spin splitting is observed near the Γ point, where a Dirac-like point emerges at the crossing of the Rashba-split bands, locally mimicking the dispersion relations of relativistic Dirac fermions and thus termed Dirac-like fermions. As well established in magnetically doped three-dimensional topological insulators[30,31], breaking TRS is an effective way to open an energy gap at the Dirac point[9,10]. Based on the band structure shown in Fig. 2d, we further introduced the effect of interfacial magnetism on the electronic band structures. To further capture the interface-induced magnetism of 2DEG arising from Eu doping and magnetic proximity effect, a tight-binding Hamiltonian was constructed using Wannier orbitals with an in-plane Zeeman term of $E_Z = 0.5$ eV included[19,32-34]. As shown in Fig. 2e, the TRS opens the Dirac points of both $L_{z,-}$ and $L_{z,+}$ orbitals at the Γ point, with the $L_{z,+}$ branch exhibiting a larger band splitting. Therefore, the spin-polarized 2DEG at the ETO/KTO (111) interface provides an ideal platform to explore the influence of a Dirac gap opening driven by TRS breaking due to ferromagnetism.

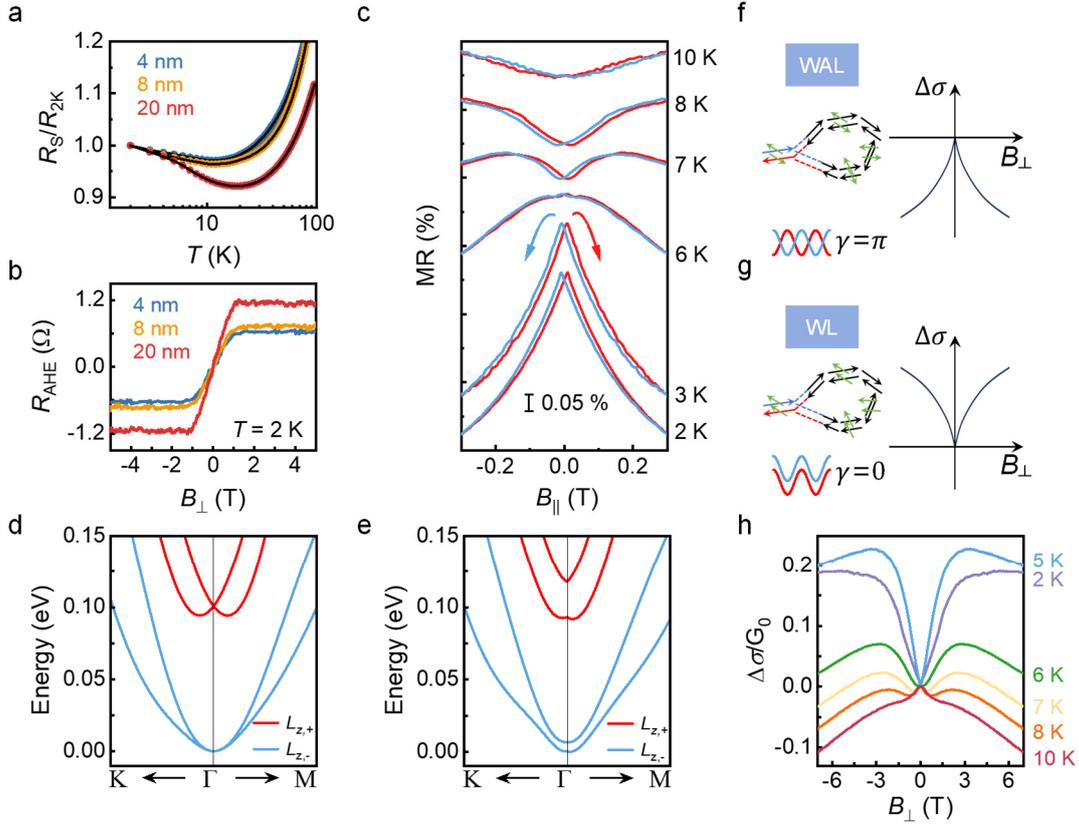

**Figure 2.** (a) Temperature dependence of the normalized sheet resistance $R_S/R_{2K}$, where $R_{2K}$ is the sheet resistance at 2 K, for ETO/KTO (111) samples with EuO layer thicknesses of 4, 8, and 20 nm. Solid lines represent the theoretical fits based on Eq. (1). (b) Anomalous Hall resistance as a function of magnetic field measured at 2 K. (c) Magnetoresistance of the ETO (20 nm)/KTO (111) sample measured at different temperatures with the in-plane magnetic field $B_\parallel$ perpendicular to applied current. For clarity, the MR-$B_\parallel$ curves at different temperatures are vertically shifted upward. (d) Electronic band structure of the KTO (111) surface obtained from first-principles calculations with SOC, showing the lower $L_{z,-}$ and upper $L_{z,+}$ orbitals. (e) Further considering the effect of magnetism, the calculated band structure shows that both the $L_{z,-}$ and $L_{z,+}$ orbitals exhibit a Dirac gap opening. Sketches of (f) WAL at a Berry phase $\gamma = \pi$ and (g) WL at a Berry phase $\gamma = 0$. (h) MC as a function of out-of-plane magnetic field measured at different temperatures.

To elucidate the impact of band structure on the quantum transport of Dirac-like fermions, we systematically studied quantum interference corrections to conductivity. Quantum interference between coherent electronic wavefunctions propagating along time-reversed, self-crossing paths gives rise to detectable quantum corrections to conductivity, manifesting as either WAL or WL[9,10]. As schematically illustrated in Fig. 2f, WAL stems from destructive interference induced by spin precession in the presence of Rashba SOC, which introduces a relative Berry phase of $\gamma = \pi$[35]. The resulting suppression of backscattering enhances conductivity near zero magnetic field. Conversely, WL arises from constructive interference between time-reversed, closed diffusive trajectories, where the accumulated Berry phase difference is $\gamma = 0$ (Fig. 2g). This enhances backscattering and consequently suppresses conductivity near zero magnetic field. In both cases, the application of an external magnetic field breaks time-reversal symmetry, suppresses phase coherence, and thereby eliminates the interference-induced quantum corrections.

Magnetoconductivity (MC) measurements under an out-of-plane magnetic field $B_\perp$ provide an effective approach to experimentally identify the WAL and WL effects[36]. Figure 2h presents the temperature-dependent MC, defined as $\Delta\sigma(B_\perp) = \sigma(B_\perp) - \sigma(0)$ and expressed in units of quantum conductance $G_0 = e^2/\pi h$, for a spin-polarized 2DEG at the ETO(20 nm)/KTO (111) interface, measured over the temperature range of 2 to 10 K. At even higher temperatures ($T > 10$ K), the MC exhibits a parabolic $B^2$-dependent background (see Fig. S10), which originates from orbital effects related ordinary MC (OMC)[37]. At $T = 10$ K, the MC is negative with a distinct cusp-like shape, a hallmark of WAL, and gradually evolves into a peak-to-shoulder feature as the temperature decreases to 7 K. Remarkably, a dramatic transition occurs between 7 and 6 K, where the MC reverses its sign and changes from a peak to a dip near zero magnetic field. At $T \leq 6$ K, the positive MC curve exhibits a distinct cusp near zero field, a characteristic feature of WL, indicating a crossover from WAL to WL as the temperature decreases. Therefore, our results reveal a temperature-dependent competition between WL and WAL of Dirac-like fermions in the spin-polarized 2DEG at the ETO/KTO (111) interface.

Next, we perform a quantitative analysis of the competing WL and WAL scattering channels of Dirac-like fermions in our 2DEG system. The observed WAL-WL crossover can be described by the formula derived in Refs. [4,5]:

$$\Delta\sigma(B) = \sum_{i=0,1} \frac{\alpha_i e^2}{\pi h}\left[\Psi\left(\frac{\ell_B^2}{\ell_{\phi i}^2}+\frac{1}{2}\right) - \ln\left(\frac{\ell_B^2}{\ell_{\phi i}^2}\right)\right] - A_K \frac{\sigma(0)}{G_0}\frac{B^2}{1+CB^2} \quad (2)$$

where $\Psi(x)$ is the digamma function, defined as $\Psi(x) = \ln(x)+\Psi(1/2+1/x)$, the prefactors $\alpha_0$ and $\alpha_1$, with opposite signs, quantify the weights of the WL and WAL contributions, respectively, and $\ell_{\phi i}$ is the corresponding phase coherence length. Note that the prefactor $\alpha_0$ for the WL channel is positive, with a limiting value of 0.5, while $\alpha_1$ for the WAL channel is negative, with a limiting value of $-0.5$[9,10]. The last term, known as the Kohler term, accounts for the orbital effect of the magnetic field and involves the parameters $A_K$ and $C$. OMC follows Kohler's rule and becomes more pronounced with increasing Hall mobility[38]. Therefore, the total MC arises from the combined contributions of WL, WAL, and OMC. In Figs. 3a and 3d, the MC data as a function of temperature are presented as a curve and a color map, respectively. As shown by the fitting curves (black solid lines) in Fig. 3a, Eq. (2) captures the key features of the experimental data, providing excellent fits to the temperature-dependent MC curves at low magnetic fields with appropriate parameters. For clarity, only a subset of representative temperatures is displayed in Fig. 3a, and the complete fitting results for all temperatures are provided in the Supplementary Fig. S10. Based on the obtained fitting parameters, the separate contributions of WL and WAL to the MC, after subtracting the Kohler background, are plotted as a function of $B_\perp$ in Fig. 3b. The temperature dependence of the fitting parameters $\alpha_0$ and $\alpha_1$ is summarized in Fig. 3c, and the remaining fitting parameters are listed in Table S3. When $T > 8$ K, the absolute value of $\alpha_0$ is consistently smaller than that of $\alpha_1$, indicating that the MC is primarily dominated by the WAL channel. As the temperature decreases to 8 K and below, the absolute value of $\alpha_0$ gradually surpasses that of $\alpha_1$, suggesting a transition in the dominant contribution to MC from WAL to WL. This competing evolution of the WL and WAL channels with decreasing temperature is also clearly observed in Fig. 3b.

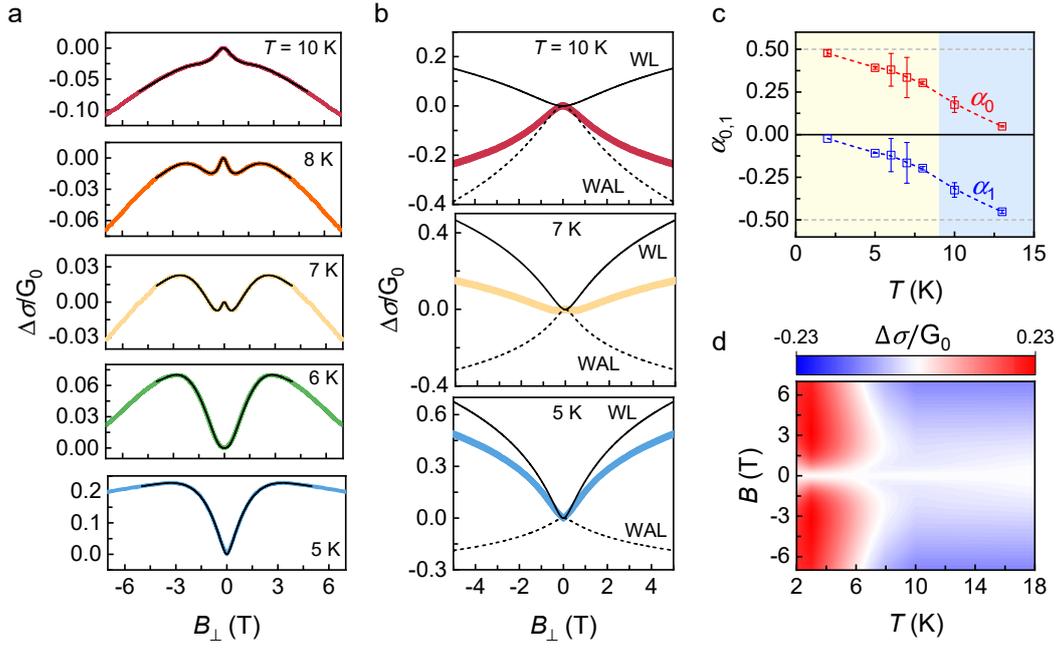

**Figure 3.** (a) Temperature-dependent MC curves and the corresponding fitting results (black solid curves) based on Eq. (2). (b) Decomposition of the MC data after subtracting the OMC component into WAL and WL contributions. (c) Temperature dependence of the fit parameters $\alpha_i$ (i = 0,1). (d) Color map of the MC data as a function of temperature.

This temperature-dependent anomalous evolution of the MC curves can be attributed to the interplay between Rashba SOC and magnetism in the 2DEG at the ETO/KTO (111) interface. The strong SOC, originating from the heavy 5$d$ element Ta, induces spin-momentum locking in the conducting two-dimensional plane, leading to the emergence of WAL, a phenomenon widely reported for KTO-based 2DEGs[1,6,15,39,40]. As shown in Fig. 2c, at temperatures above 8 K, the MR exhibits no hysteresis, indicating that the system is in a nonmagnetic regime. In this case, the band structure corresponds to that shown in Fig. 2d, where Rashba SOC acts as the dominant mechanism, leading to the WAL-type MC. However, upon cooling below the ferromagnetic transition temperature (~8 K), hysteretic MR behavior emerges (Fig. 2c), indicating that the 2DEG enters a ferromagnetic state. The long-range magnetic ordering breaks TRS and opens a magnetic gap at the Dirac-like point, corresponding to the band structure shown in Fig. 2e. In this regime, the opening of a magnetic gap induces a WL scattering channel, resulting in a crossover to WL-dominated MC. This behavior is analogous to that observed in magnetically doped topological insulators[9-13]. Our results highlight the crucial role of the ferromagnetism-induced energy gap in modulating the competition between quantum interference channels in the Dirac-like 2DEG system.

**3. Optical gating of quantum interference in Dirac-like Fermions**

Optical gating provides an effective means to modulate carrier density, band occupation, and spin-orbit interaction in low-dimensional electronic systems[1,15,39]. The transport properties of Dirac-like fermions in oxide 2DEGs are expected to be highly sensitive to photoinduced variations in band filling. To investigate this effect, the photoresponse of the ETO (20 nm)/KTO (111) sample was measured under light illumination, as depicted in the inset of Fig. 4a. The $R_S$-$T$ curves recorded before and after light illumination (light power $P$ = 1 mW and wavelength $\lambda$ = 405 nm) are displayed in Fig. 4a. Notably, upon light illumination, the Kondo effect is completely suppressed, accompanied by a sizable decrease in $R_S$ below 50 K, consistent with previous reports in oxide heterostructures such as a-LAO/KTO[15] and CaZrO$_3$/KTO[39]. Figure 4b shows the Hall resistance $R_{xy}$ as a function of the out-of-plane magnetic field $B_\perp$ at 5 K under illumination with varying $P$. In the dark, $R_{xy}$ varies linearly with $B_\perp$. As $P$ increases, a curvature in the $R_{xy}$-$B_\perp$ relationship appears at $P$ = 0.15 mW and becomes increasingly pronounced with higher light intensity. We analyze the Hall effect results in combination with the calculated band structure of the KTO (111) surface shown in Fig. 2. Under dark condition, a linear $R_{xy}$-$B_\perp$ curve is observed, indicating that only one type of charge carriers exists in the system, occupying the lower-energy $L_{z,-}$ subbands. Under illumination, the Hall effect exhibits nonlinear behavior, with the hump feature becoming more prominent as the light power increases. As depicted by the black solid fitting curves in Fig. 4b, this nonlinear Hall effect can be satisfactorily described by the combined two-band model and AHE, as detailed in Supplementary Note S2. In addition, the anomalous Hall resistance under different light power is shown in Fig. S11. The above results indicate the presence of a second type of charge carriers. Since the photon energy employed in our experiment (~3.1 eV) is lower than the band gap of KTO (~3.58 eV), light illumination promotes electrons from in-gap states into the conduction band of KTO[1,15,39,41]. This photoinduced carrier pumping increases the carrier density, shifting the Fermi level $E_F$ upward to cross the bottom of the higher-lying $L_{z,+}$ subbands, thereby giving rise to a second species of charge carriers. The extracted carrier density and the corresponding Hall mobility as functions of laser power $P$ are presented in Figs. 4c and 4d, respectively. The carrier density of the $L_{z,-}$ electrons ($n_{S1}$~1.85×10$^{14}$ cm$^{-2}$), is nearly independent of $P$, with a corresponding Hall mobility $\mu_1$ of ~22 cm$^2$ V$^{-1}$ s$^{-1}$. As $P$ increases from 0.15 to 1 mW, the carrier density of the $L_{z,+}$ electrons ($n_{S2}$) increases from 3.0×10$^{10}$ to 5.9×10$^{12}$ cm$^{-2}$, an enhancement of nearly 2 order of magnitude, yet it remains much smaller than $n_{S1}$. The corresponding Hall mobility $\mu_2$ rises from 2663 to 8388 cm$^2$ V$^{-1}$ s$^{-1}$, exceeding $\mu_1$ by more than two orders of magnitude. Obviously, photoexcitation alters the occupation of electronic states by shifting the $E_F$ upwards, and the emergence of a second type of high-mobility charge carriers results in a drop in the $R_S$-$T$ curve and the suppression of the Kondo effect.

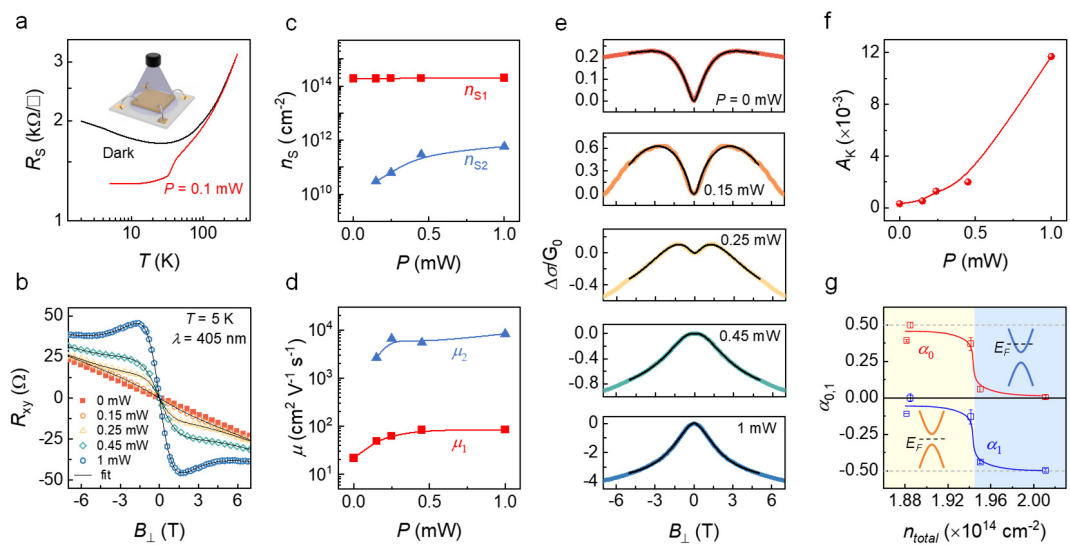

**Figure 4.** (a) Temperature dependence of sheet resistance at the ETO/KTO (111) interfaces, measured with and without light illumination (light power has been set to 0.1 mW). Inset is a sketch for experiment setup. (b) Hall resistance as a function of magnetic field, measured at 5 K by fixing light power to a series of constant values between 0 and 1 mW. (c, d) Extracted carrier density and Hall mobility, respectively, as functions of laser power, obtained from fits to the experimental data in (b). (e) MC data measured at 5 K under different light powers. Black lines are the best fits using Eq. (2). (f) Kohler parameter describing the orbital effect of a perpendicular magnetic field as a function of $P$. (g) The evolution of fit parameters $\alpha_i$ (i = 0,1) fitted by Eq. (2) as a function of carrier density.

We further measured the MC under illumination with varying $P$ from 0 to 1 mW at $T = 5$ K, as shown in Fig. 4e. Strikingly, an optically controlled quantum interferences competition between WAL and WL effect is observed, and the fitting curves (black solid lines) based on Eq. (2) exhibit excellent agreement with the experimental data. As $P$ increases, the MC evolves from WL-typed behavior to WAL-typed behavior superimposed on a typical two-carrier-induced OMC background. With the photoexcitation of high-mobility carriers, the contribution from the orbital effect becomes increasingly significant, as evidenced by the increase of the Kohler parameter $A_K$ with $P$ (Fig. 4f). To explore the influence of the band filling tuned via optical gating, the fitting parameters $\alpha_0$ and $\alpha_1$ are plotted as functions of $n_{total}$ in Fig. 4g, where $n_{total} = n_{S1}$ when only one type of charge carrier is present and $n_{total} = n_{S1}+n_{S2}$ when two types of charge carriers coexist. As $n_{total}$ increases, $\alpha_0$ decreases while $|\alpha_1|$ increases, indicating that the WL contribution is significantly suppressed, whereas the WAL contribution becomes more pronounced. Noting that a magnetic gap opens at the Dirac-like point when $T = 5$ K, the observed crossover from WL to WAL with increasing $n_{total}$ can be attributed to the tuning of the $E_F$ position relative to the magnetic gap[9,10]. When $E_F$ is located inside the magnetic gap, the MC exhibits WL-dominated behavior, with the Berry phase $\gamma$ tending toward 0. Upon illumination, photoinduced carrier pumping elevates $E_F$ into the higher-lying $L_{z,+}$ subbands and $E_F$ moves to intersect the Dirac-like cone. Consequently, the Berry phase $\gamma$ shifts toward $\pi$, and the WAL contribution becomes dominant. The schematic of the gapped Dirac-like band structure, along with the corresponding $E_F$ position, is shown in the inset of Fig. 4g. Therefore, optical gating can effectively modulate the WL-WAL competition via the Fermi-level-dependent Berry phase, providing a novel route for controlling quantum interference phenomena.

In summary, we have comprehensively investigated the quantum interference effects in spin-polarized 2DEGs at the interface of ferromagnetic ETO films and KTO (111) substrates. The presence of long-range ferromagnetic order in the 2DEGs is evidenced by the AHE and hysteretic MR, arising from both interfacial Eu doping and magnetic proximity effect. The temperature-dependent competition between WAL and WL in quantum corrections to MC is observed, driven by the breaking of time-reversal symmetry due to ferromagnetism and the opening of a magnetic gap at the Dirac-like point. Additionally, optical gating can effectively modulate this competitive behavior by tuning the Fermi level, highlighting the significant role of band filling in governing the behavior of Dirac-like fermions at the ETO/KTO (111) interface. These findings not only elucidate the intricate interplay among spin polarization, quantum interference, and band topology in oxide interfaces, but also offer a pathway towards tunable quantum transport through optical gating in oxide-based heterostructures.

# METHODS
## Sample fabrication and characterization

ETO films were grown on (111)-oriented KTO substrates by the technique of pulsed laser deposition. A KrF Excimer laser (wavelength is 248 nm) was employed. The repetition rate was 2 Hz and the fluence was ~2 J/cm$^2$. During film growth, the substrate temperature was kept at 700 °C and the oxygen pressure was set to $1\times10^{-4}$ Pa. After deposition, the temperature of the sample was furnace-cooled to room temperature under the same atmosphere. Film thickness was determined by the number of laser pulses, which has been carefully calibrated by small-angle X-ray reflectivity. The surface morphology of the samples was evaluated via atomic force microscopy (AFM, MultiMode 8, Bruker). The crystal structure was determined by high-resolution X-ray diffractometry (XRD, D8 Discover, Bruker) with Cu-Kα radiation. Lattice images were recorded by a high-resolution scanning transmission electron microscope (STEM) with double C$_S$ correctors (JEOL-ARM200F).

## Magnetic and Transport Measurements

Magnetic properties were measured by a Quantum-designed superconducting quantum interference device vibrating sample magnetometer (SQUID-VSM). The electrical transport measurements were performed on a Quantum-designed physical properties measurement system (PPMS). Ultrasonic wire bonding (Al wires of 20 μm in diameter) was used for electric connection. An applied current of 10 μA was used. The van der Pauw geometry was adopted to measure the temperature dependence of sheet resistance, Hall effect, and out-of-plane magnetoresistance measurements, and the standard four-probe technique was employed for in-plane magnetoresistance measurements. To investigate the effect of photoexcitation on transport behavior, semiconductor laser beam (λ = 405 nm) was introduced into PPMS by an optical fiber to illuminate the samples. The laser spot size was ~2 mm in diameter. The laser power (0~1 mW) and the corresponding photon flux (0~6×10$^{16}$ cm$^{-2}$ s$^{-1}$) were calibrated at the output of the optical fiber. All light-controlled measurements were performed after a sufficient waiting period to allow the resistance to stabilize under illumination.

## First-principles calculation

Electronic structure calculations were carried out within the framework of DFT using the projector augmented wave method[42], as implemented the Vienna *ab initio* simulation package[43,44]. The exchange-correlation potential was described by generalized gradient approximation (GGA) of the Perdew-Burke-Ernzerhof (PBE) functional[45]. A slab model consisting of six unit cells of KTO along the [111] direction was constructed. The Brillouin zone was sampled using a $\Gamma$-centered ($6 \times 6 \times 1$) $k$-mesh and a plane-wave energy cutoff of 500 eV was employed. To capture the correlation effects of Ta $5d$ electrons, the rotationally invariant DFT+U scheme was adopted with $U_{\text{eff}} = 4.0$ eV[39,46,47]. Given the strong relativistic nature of Ta, SOC was also included in the band structure calculations. The tight-binding model Hamiltonian was constructed with the Ta-$5d$ and O-$2p$ orbitals using the Wannier90 package[32,48]. To incorporate the effect of interfacial ferromagnetism, an effective Zeeman term was further introduced using the WannierTools package[33,34,49].

**Data availability**

The authors declare that data generated in this study are provided in the paper and the Supplementary Information file. Further datasets are available from the corresponding author upon request.

**Acknowledgments**

This work has been supported by the Science Center of the National Science Foundation of China (Grant No. 52088101), the National Key Research and Development Program of China (Grant No. 2022YFA1403302, No. 2024YFA1410200, No. 2021YFA1400300, No. 2021YFB3501200, No. 2021YFB3501202, and No. 2023YFA1406003), the National Natural Science Foundation of China (Grant No. 12474103, No. T2394470, No. T2394474, No. 12274443, No. 92263202, No. U23A20550, and No. 22361132534), the Strategic Priority Research Program B of the Chinese Academy of Sciences (Grant No. XDB33030200), and the Beijing Outstanding Young Scientist Program. The authors acknowledge the Center for Micro-Nano Innovation of Beihang University for the sample morphological characterization.

**Author contributions**

H.Z. conceived the project and proposed the strategy. H.Z., J.R.S., and W.S.Z. supervised the study. D.M.T. and L.C. prepared the samples. D.M.T., M.L., and Y.T.B. developed the experimental setup and performed the transport measurements. X.B.C. carried out the theoretical calculations. H.Z., D.M.T., X.B.C., and J.R.S. analyzed the experimental and theoretical data. F.X.H., and B.G.S. contributed to data analysis and discussion. H.Z. and D.M.T. wrote the manuscript with input from all authors. All authors contributed to manuscript revision.

**Conflict of Interest**
The authors declare no conflict of interest.